\documentclass[a4paper,14pt]{extarticle}
\usepackage{cmap} 
\usepackage[cp1251]{inputenc}
\usepackage[english,russian]{babel}

\usepackage{braket,amssymb,amsmath,array,commath,mathtext,wrapfig,tikz}
\usepackage[arrows=pgf-filled,version=3]{mhchem}
\usepackage{bm}
\usepackage{amsthm,amsfonts,amscd}
\usepackage{graphicx}
\usepackage{cite}
\usepackage{lscape}
\usepackage{geometry}
\usepackage{indentfirst}
\usepackage{afterpage,soul,ulem,dcolumn,changebar,longtable,hhline,multirow,makecell}
\usepackage[small,bf]{caption}
\makeatletter
\renewcommand{\@biblabel}[1]{#1.\hfil} 
\makeatother

\begin{document}
\renewcommand{\refname}{References}

	\begin{center}
\textit{Geodesics for particles with negative energy in Kerr's  metric.}  
	\end{center}
	
	\begin{center}
\textbf{V. D. Vertogradov} \\
	\end{center}
	\begin{center}
		Physics department, Herzen state Pedagogical University of Russia,
		
		48 Moika Emb., Saint Petersburg 191186, Russia
\\
\text{email}  vdvertogradov@gmail.com
	\end{center}

\textbf{Abstract} \\

According to Penrose’s effect, particles with negative energy can exist in the ergospheres of rotating black holes. We analyze geodesics for such particles and show that there are no circular and elliptic orbits in the ergosphere of a rotating black hole. We also show that there are geodesics which begin and terminate at the singularity and present the conditions under which such geodesics do not begin and terminate at the singularity.
 
\textbf{key words:} black holes, Kerr metric, geodesics, negative energy. \\
\textbf{PACS numbers:} 04.70.-s, 04.70.Bw, 97.60.Lf \\

\section{1. Introduction}

In 1969, Roger Penrose predicted an effect~\cite{bib:1} which was later called Penrose’s mechanism. According to this effect, there are particles with negative energy in the ergospheres of rotating black holes. In this regard there is a question on the description of geodesics for such particles. Then, Contopoulos~\cite{bib:2} showed that such geodesics achieve the event horizon in finite proper time in the case of the equatorial plane. However, there is a question on the beginning of such geodesics.

Our spacetime must be geodesically complete. In other words, all geodesics must begin either at a singularity or at infinity. As geodesics for particles with negative energy cannot overcome the static limit, they either begin at the black hole singularity or have a closed orbit in the ergosphere, and the proper time of their motion in the ergosphere must be infinite in spite of the fact that at some moment this geodesic moves inside the gravitational radius.

In this paper we will prove the absence of circular and elliptic orbits (under elliptic orbits we mean closed, noncircular bounded lines) in the ergosphere and also give the condition of the end and the beginning of such geodesics at the singularity of the black hole. The proof given in this article differs from the one given by Grib and Pavlov~\cite{bib:3, bib:5} . Also, here we will prove that the orbits which do not begin at the singularity are either circular or elliptic and also stable inside the Cauchy horizon. This has not been considered before.

It is worth emphasizing that geodesics for particles with negative energy are white hole solutions~\cite{bib:6}. Such geodesics appear in the ergosphere from a region inside the gravitational radius. It follows therefrom that the cosmic censorship principle is violated. (Of course if we consider the real case of a collapsing star but not an eternal black hole, then sometimes the cosmic censorship principle can be valid, but in this case one needs an explanation of the origin of such geodesics for particles with negative energy.) According to this principle, we can never get any information from a region inside the gravitational radius of a black hole, but we show that it is not so. It means that the particle can move along such geodesics and can bring information from a region inside the gravitational radius. The cosmic censorship principle can be understood as the absence of white holes in the nature. However, as we show, “white hole” geodesics must always exist in the region outside the gravitational radius.

The system of units $G = c = 1$ will be used.

\section{2. Absence of circular and elliptical orbits in the ergosphere}
\setcounter{equation}{0}

Neutral, rotating black holes are described by the Kerr metric which is given in Boyer-Lindquist coordinates by~\cite{bib:1}

	\begin{equation}
	ds^2=\frac{\Delta}{\rho^2}(dt-a\sin^2\theta d\varphi)^2-\frac{\sin^2\theta}{\rho^2}\left[(r^2+a^2)d\varphi-adt\right]^2-\frac{\rho^2}{\Delta}(dr)^2-\rho^2(d\theta)^2\,,
	\end{equation}

where:
\begin{equation}\Delta=r^2-2Mr+a^2\,,\end{equation}
	\begin{equation}\rho^2=r^2+a^2\cos^2\theta\,,\end{equation}
$M$- mass of a black hole, $a$ - its angular momentum.
It's more convenient to change this coordinates on dimensionless ones:

	\begin{equation}x=\frac{r}{M}, A=\frac{a}{M}\,.\end{equation}
Then,

	\begin{equation}\rho_x=x^2+A^2\cos^2\theta, \Delta_x=x^2-2x+A^2\,.\end{equation}

The event horizon and the Cauchy horizon are denoted by $x_H$ and $x_C$, respectively:
	\begin{equation}x_H=1+\sqrt{1-A^2}\,,\end{equation}
	\begin{equation}x_C=1-\sqrt{1-A^2}\,.\end{equation}
Geodesics in Kerr's metric are given by~\cite{bib:1}:

	\begin{equation}\label{1.4} \varepsilon^{-2}\rho^4_x\left(\frac{dx}{d\tau}\right)^2=x^4+(A^2-\xi^2-\nu)x^2+2x\left[\nu+(A-\xi)^2\right]-A^2\nu-\varepsilon^{-2}x^2\Delta_x\,,\end{equation}

	\begin{equation}\label{1.9}\rho^2_x\frac{dt}{d\tau}=\left(\frac{\varepsilon}{\Delta_x}\right)\left[(x^2+A^2)^2-A^2\Delta_x\sin^2\theta-2A^2x\xi\right]\,,\end{equation}
where  - $\tau$ the proper time of a particle, $\varepsilon$  - specific energy of a particle which takes only negative value, $\xi=\frac{L_z}{\varepsilon}, \nu=\frac{Q}{\varepsilon^2}$  - impact parameters,  $L_z$ - projection component of angular momentum on spinning axis which takes negative values and $Q$ - Carter's constant.
The main condition is forward motion  in time.  In other words $\frac{dt}{d\tau}>0$.
Effective potential is given by:
	\begin{equation} \label{1.8}V_{eff}+\left(\frac{dx}{d\tau}\right)^2=0\,.\end{equation}
Then from  \ref{1.4} and \ref{1.8} we find:

	\begin{equation}\label{1.11}V_{eff}=-\frac{\varepsilon^2}{\rho^4_x}\left\{x^4+(A^2-\xi^2-\nu)x^2+2x\left[\nu+(\xi-A)^2\right]-A^2\nu-\varepsilon^{-2}x^2\Delta_x\right\}\,.\end{equation}

The condition of existence of circular orbits is~\cite{bib:4}:
\begin{equation}V_{eff}(x_0)=0, \frac{dV_{eff}(x)}{dx}|_{x=x_0}=0\,.\end{equation}
Putting the expression \ref{1.11} and its  derivate to zero and solving the system  of equations relatively  on impact parameters we found values of impact parameters $\xi, \nu$for  which circular orbits exist:

	\begin{equation}\label{1.13}\xi_{\pm}=A^{-1}(x-1)^{-1}(x^2-A^2\pm x\Delta_x\sqrt{1-\varepsilon^{-2}(1-x^{-1})})\,.\end{equation}
Values of $\nu$ are too complicated and as we  shall not use them further, so its values aren't given in this article. To prove that circular orbits don't exist one needs to put the expression \ref{1.13} into \ref{1.9} . If we prove that $(x^2+A^2)^2-A^2\Delta_x\sin^2\theta-2Ax\xi>0$ putting  $\theta=\frac{\pi}{2}, \xi=\xi_+, \sqrt{1-\varepsilon^{-2}(1-x^{-1})}=1$, then we'll prove this inequality for all $\theta, \xi\leq\xi_+, \sqrt{1-\varepsilon^{-2}(1-x^{-1})}\leq 1$. Putting all this into expression \ref{1.9}  and doing simple calculations we find:

	\begin{equation}\rho^2_x\frac{dt}{d\tau}\leq \left(\frac{\varepsilon}{\Delta_x}\right)(x^2\Delta_x)\,.\end{equation}
As $\Delta_x>0$ in the ergosphere and $\varepsilon<0$ we get $\frac{dt}{d\tau}<0$ and the condition of forward motion in time is broken. It follows therefrom that there aren't circular orbits  in the ergosphere. 

The condition of existence of elliptical orbits ( the difinition is in introduction) is~\cite{bib:4}:
	\begin{equation}V_{eff}(x_0)=0, \frac{dV_{eff}(x)}{dx}|_{x=x_0}>0\,.\end{equation}
This is true if impact parameters $\xi$ belong to the interval $(\xi_-,\xi_+)$ but above it was found  that for all $\xi\leq \xi_+$ the condition of forward motion  in time is broken so elliptical orbits also don't exist. 

It follows therefrom that particles with negative energy achieve the event horizon in finite proper time.

So because circular and elliptical orbits don't exist in the ergosphere and such geodesics can't  overcome  the static limit, we come to the conclusion that such geodesics emerge in the ergosphere from  a region inside the gravitational radius.

\section{3. The conditions of the terminating of geodesics in the singularity}
\setcounter{equation}{0}

Now, we consider conditions of the terminating of the geodesics in the singularity. If the particle is Crossing the event horizon  it achieves the Cauchy's horizon. 

Here we consider the simplest  case  of massless geodesics in the equatorial plane. In this case the effective potential is given by:
	\begin{equation}V_{eff}=-\frac{1}{2}\left[\varepsilon^2+\frac{2}{x^3}(A\varepsilon-L_{z})^2+\frac{A^2\varepsilon^2-L_z^2}{x^2}\right]\,.\end{equation}

Here geodesics are given by:
\begin{equation}
(\frac{dx}{d\tau})^2=2V_{eff} \,,
\end{equation}
\begin{equation} \label{mgt}
\frac{dt}{d\tau}=\frac{1}{\Delta_x}\left [(x^3+A^2x+2A^2)\varepsilon-2L\right ]\,.
\end{equation}

Geodesic  terminates in the singularity $x=0$ if  the effective potential in the interval $(0,x_C)$ takes negative values. Notice that it's enough to prove that:
	\begin{equation}\alpha(x)=\varepsilon^2x^3+(A^2\varepsilon^2-L_z^2)x+2(A\varepsilon-L_z)^2>0\,,\end{equation}
	in the interval $(0,1)$.

In the beginning we calculate values $\alpha$ in points $0,1$:
	\begin{equation}\alpha(0)=2(A\varepsilon-L_z)^2>0\,,\end{equation}

	\begin{equation}\alpha(1)=L_z^2-4A\varepsilon L_z+\varepsilon^2+3A^2\varepsilon^2 >0\,.\end{equation}

The inequality is true because $A<1$  and solving $\alpha(1)$ relatively to $L_z$ we find that the discriminant is less then zero. 

Now we show that the function $\alpha(x)$ has no extremes in the interval $(0,1)$. If we show it we'll prove that the function $\alpha(x)$ is positive in this interval.

	\begin{equation}\frac{d\alpha(x)}{dx}=3\varepsilon^2x^2+A^2\varepsilon^2-L_z^2\,.\end{equation}

Putting this expression to zero we find:
	\begin{equation}x=\pm\sqrt{\frac{L^2_z-A^2\varepsilon^2}{3\varepsilon^2}}\,.\end{equation}

The solution with negative sign doesn't suit us. Next we prove that:
	\begin{equation}\frac{L_z^2-A^2\varepsilon^2}{3\varepsilon^2}>1\,.\end{equation}

It's true because the parameter $0<A<1$ and the condition  of forward motion in time demands $L^2_z\geq 4A^2\varepsilon^2$. To prove that $|L_z|\geq |2\varepsilon|\geq |2A\varepsilon|$ we need to put $|L_z|=2|\varepsilon|$ into \ref{mgt} then doing simple calculation  we find that  if $"|L_z|\leq 2|\varepsilon|$ then  the condition of forward motion in time is broken.

We've proved  that the function $\alpha$ has no  extremes in the interval $(0,1)$ and as it is positive in points $0,1$ so the function decrease in the interval $(0,1)$.  It follows therefrom  that all massless geodesics terminate  in the singularity. 

The case of timelike geodesics is more difficult and here we  give only results~\cite{bib:5}. 
All geodesics for particles with energy  which belongs to the interval $(-\infty, -\frac{1}{3}]$ terminate and begin in the singularity.
If an energy belongs to the interval 
\begin{equation}\left(-\frac{1}{3},0\right)\end{equation}
 and  the inequality: 	
\begin{equation}A>4\varepsilon(1-\lambda)\end{equation}
 is true, where $\lambda=\frac{L_z}{A\varepsilon}>2$  is real number, geodesics for such particles don't terminate and begin in the singularity.

\section{4. Geodesics which don't terminate  and begin in the singularity inside the Cauchy horizon}
\setcounter{equation}{0}

Now we consider timelike geodesics which don't terminate and begin in the singularity in the equatorial  plane inside the Cauchy horizon. In this case effective potential is given by:
	
\begin{equation}
V_{eff}=-\frac{1}{2}\left[\varepsilon^2+\frac{2}{x^3}(L_z-A\varepsilon)^2+\frac{A^2\varepsilon^2-L^2-\Delta_x}{x^2}\right]\,.
\end{equation}
The condition of the stability of orbits is given by ~\cite{bib:4} $\frac{d^2V_{eff}(x)}{dx^2}|_{x=x_0}<0, V_{eff}(x_0)=0$. In the beginning we'll show that such orbits are either circular or elliptical. It's enough to prove that $\alpha(x_0)=0, \frac{d\alpha(x)}{dx}|_{x=x_0}\leq 0$ where:
\begin{equation}\alpha(x)=\beta x^3+2x^2+(A^2\beta-L_z^2)x+2(A\varepsilon-L_z)^2\,.\end{equation}
here $\beta=\varepsilon^2-1<0$,
\begin{equation}\frac{d\alpha(x)}{dx}=3\beta x^2+4x+A^2\beta-L^2\,.\end{equation}
Now we find roots of this equation
\begin{equation}
x_{1,2}=\frac{-2\mp \sqrt{4-3\beta(A^2\beta-L_z^2)}}{3\alpha}
\,.\end{equation}
The first derivative is less then zero in the interval $(x_1,x_2)$. It's worth noticing that the function $\alpha(x)$  is growing in the interval $(x_1,x_2)$. Also if values of the energy and of the angular momentum are very small  according to conditions above  then $\alpha(x_2)>0, \alpha(x_1)<0$. Then there is the point $x_0$ in the interval $(x_1,x_2)$ that $\alpha(x_0)=0$. It follows therefrom that such orbits are either circular or elliptical. 

Next we prove the stability of such orbits. The condition of stability of orbits in case of the function $\alpha(x)$ is $\alpha(x_0)=0, \frac{d^2\alpha(x_0)}{dx^2}>0$.
We find orbits are stable if $x<\frac{2}{3\beta}$. But in this point, the function is positive and it follows therefrom that $\frac{d^2\alpha(x_)}{dx^2}<0$ and such orbits are stable.

\section{5. Conclusion}

We have shown that circular and elliptic orbits with negative energy in the ergosphere of a rotating black hole do not exist. We have also found that such orbits which do not begin and terminate at the singularity are either elliptic or circular and stable inside the Cauchy horizon. We have also obtained the condition of the end and the beginning of geodesics at the singularity. Since they begin at the singularity and appear in the ergosphere from under the event horizon, the cosmic censorship principle is violated. It means that we can get information from under the event horizon. It is worth emphasizing that the cosmic censorship principle is violated in the case of an eternal black hole while at real collapse it can happen that due to the absence of a white hole and a naked singularity the cosmic censorship principle can be true. However, see ~\cite{bib:7} with an indication how it can be violated.

Acknowledgments. 
The author is grateful to A.A. Grib and Yu.V. Pavlov for helpful discussions and to the Dynasty Foundation for financial support.

\end{document}